# scientific reports

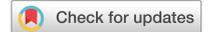

OPEN

# Robotic agricultural instrument for automated extraction of nematode cysts and eggs from soil to improve integrated pest management

Christopher M. Legner[1], Gregory L. Tylka[2] & Santosh Pandey[1]✉

Soybeans are an important crop for global food security. Every year, soybean yields are reduced by numerous soybean diseases, particularly the soybean cyst nematode (SCN). It is difficult to visually identify the presence of SCN in the field, let alone its population densities or numbers, as there are no obvious aboveground disease symptoms. The only definitive way to assess SCN population densities is to directly extract the SCN cysts from soil and then extract the eggs from cysts and count them. Extraction is typically conducted in commercial soil analysis laboratories and university plant diagnostic clinics and involves repeated steps of sieving, washing, collecting, grinding, and cleaning. Here we present a robotic instrument to reproduce and automate the functions of the conventional methods to extract nematode cysts from soil and subsequently extract eggs from the recovered nematode cysts. We incorporated mechanisms to actuate the stage system, manipulate positions of individual sieves using the gripper, recover cysts and cyst-sized objects from soil suspended in water, and grind the cysts to release their eggs. All system functions are controlled and operated by a touchscreen interface software. The performance of the robotic instrument is evaluated using soil samples infested with SCN from two farms at different locations and results were comparable to the conventional technique. Our new technology brings the benefits of automation to SCN soil diagnostics, a step towards long-term integrated pest management of this serious soybean pest.

Soybeans (*Glycine max*) are a critically important crop to the economy of the United States, the world's leading soybean producer[1–4]. The majority of soybeans produced worldwide are processed and consumed as edible oil and livestock feeds[5]. Every year, significant yield losses of soybeans are caused by a myriad of insect pests and diseases caused by viral, bacterial, fungal and nematode pathogens[6,7]. Amongst all pathogen-caused soybean diseases, the soybean cyst nematode (SCN), *Heterodera glycines*, has been ranked as the most damaging in the United States for the past two decades and yield losses caused by SCN have become a worldwide issue[8–10]. This nematode has been found in most of the soybean-producing areas of the United States and Canada, causing over a billion dollars in soybean yield loss annually with compromised seed quality and quantity[7,9,11]. A new generation of SCN can occur every 24 days during conducive, summer weather conditions, and the SCN eggs can remain dormant in the soil for over a decade in the absence of a host soybean crop[12,13]. Upon hatching from their eggs, the infective juveniles of SCN migrate to the host roots, penetrate the vascular root tissues, and establish feeding sites (called syncytia) to siphon the plant nutrients for its growing phase. Eventually the juveniles develop into swollen adult SCN females that are exposed on the root surface. After being mated by SCN males, the lemon-shaped adult SCN females produce about 50 eggs on their posterior end and subsequently fill up with 250 or more eggs internally. When egg production ceases, the females die and their body walls become hardened to form cysts around the eggs. The egg-filled cysts are the primary form in which the nematode exists in soil.

The SCN cannot be eradicated from soil, but it can be managed or controlled[12,14]. The critical start to successful SCN management lies in identifying which fields are infested with the pest and then quantifying the population density of the nematode. However, many farmers are unaware of the presence of the nematode because it often causes no obvious aboveground symptoms during the growing season[7,9,15]. This makes it nearly impossible to predict SCN population densities by visual inspection of crops or soil.

[1]Department of Electrical and Computer Engineering, Iowa State University, Ames, IA, USA. [2]Department of Plant Pathology and Microbiology, Iowa State University, Ames, IA, USA. ✉email: pandey@iastate.edu

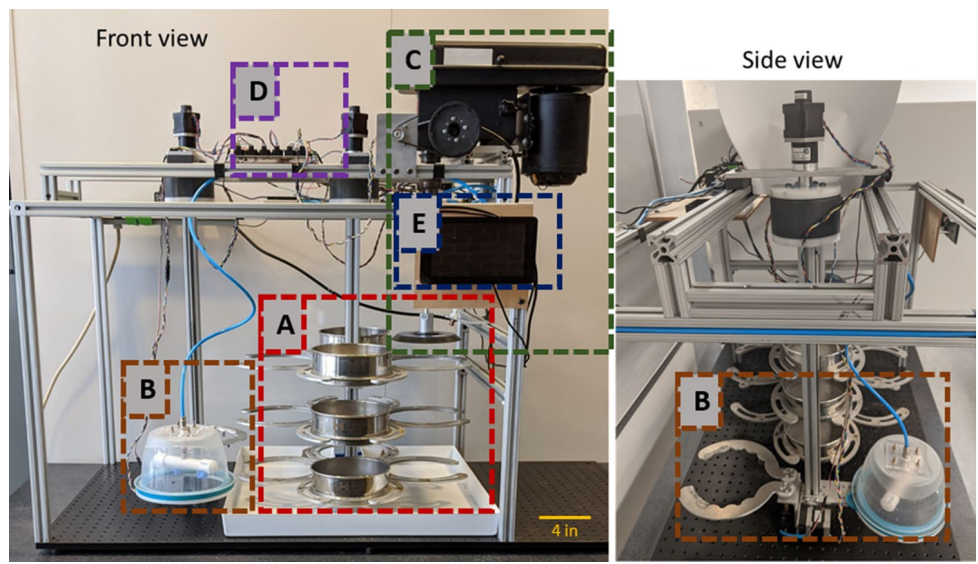

**Figure 1.** System-level overview of the robotic instrument. The instrument comprises (**A**) a stage system to hold and rotationally position sieves, (**B**) a gripper/washing system to manipulate sieves and rinse soil samples, (**C**) a grinding system to rupture cysts and release their eggs, (**D**) a control electronics board to actuate motors/sensors, and (**E**) a user interface software with a touchscreen to initiate operation modules.

Currently, the only definitive way to accurately assess SCN population densities in a field is to directly extract nematode cysts from soil and then extract and count the eggs from the cysts[12,14–18]. The techniques to extract nematode cysts from soil and eggs from cysts generally require repeated steps of mixing, washing, collecting, grinding, cleaning, and counting[19]. During the standard wet-sieving technique (Fig. S1), the soil sample is mixed with a fixed volume of water, allowed to stand by for a brief time, and then decanted multiple sieves of different pore sizes[10,20–23]. The top coarse-mesh sieve (#20 with 850-μm-diameter pores) separates the larger-sized debris from the soil–water suspension while a finer-mesh sieve (#60 with 250-μm-diameter pores) placed below it captures the egg-filled cysts[19]. A grinder (e.g. plunger made of a rubber stopper and driven by an electric motor) is used to gently break open the cyst walls on the #60 sieve, thereby releasing the eggs which are collected on an even finer sieve (#500 with 25-μm-diameter pores)[22]. A nematode-counting slide or microfluidic flow chip is then used to view and count the nematode cysts or eggs[24].

We aim to address three important concerns in the area of soil diagnostics for cyst nematodes that have direct implications to the pest management strategies practiced in the field. Firstly, manual handling is involved in multiple steps of soil processing (i.e. sieving, collecting, washing, grinding, and counting) that are laborious, time-consuming, and prone to inter-operator error and thus inconsistent analysis[24]. Secondly, soil diagnostics for cyst nematodes currently are done in specialized laboratories with trained personnel, and the time between submitting a sample to receiving the results can be more than a few weeks[25,26]. Thirdly, the average cost for extraction and quantification is expensive (approximately $25 to $50 USD per soil sample) which prevents frequent and intensive sampling of multiple points in a field[27–30]. Because of these bottlenecks, a majority of soybean-producing farmers remain ill-informed about the population densities of SCN that are resident in their fields[31].

In this work, we constructed a robotic instrument to automate the steps of extracting nematode cysts from soil and eggs from the cysts. The instrument is designed to mimic the mechanical functions involved in the manual wet-sieving extraction technique that is most commonly used in the United States (Fig. S1). A description of each component of the instrument is presented, along with its working principle and actuation method. The effectiveness of the new cyst and egg extraction instrument is evaluated by determining and comparing the population densities of SCN in soil collected from two different infested fields using the robotic instrument and wet-sieving technique. This is followed by a discussion on the technical merits of the developed technology and its value in integrated SCN management.

## Results

A system-level overview of the robotic instrument is shown in Fig. 1. The key components of the instrument are the stage system, gripper/washing system, grinding system, control electronics, and user interface software.

**Individual systems of the robotic instrument.** The 'stage system' consists of three vertical stage levels (i.e. top, middle, and bottom) as shown in Figs. 1A and 2. The top level is affixed at a set height (11.5-inch above the base). The middle level passively rests (7-inch above the base) above the bottom level (2.5-inch above the base). An actuation mechanism allows the bottom and middle levels to be raised such that sieves in the three levels can be vertically stacked (Fig. 2B–D). This automatic alignment of the sieves ensures suspended cysts and other particles and water flow directly from one sieve into another. In addition, each level consists of four sieve slots, radially affixed to a central post, to accommodate up to four sieves (Fig. 2A). Each sieve slot is designed as two semicircular fingers with recessed grooves to hold standard sieves (6-inch diameter). A geared stepper motor drives the rotational motion of the stage levels in 90-degree increments (Movie S1), while a standard stepper motor drives the linear actuation of the stage levels to vertically raise or lower the bottom and middle levels (Movie S2).

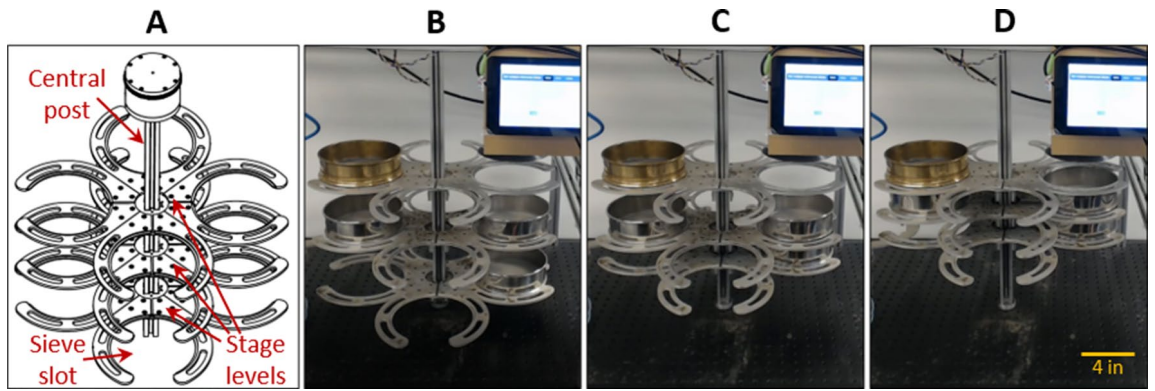

**Figure 2.** The stage system. (**A**) CAD schematic showing the three stage levels where each level has four sieve slots. (**B**) Image of the stage system in its fully uncompressed state. (**C**) Image of the stage system in its partially compressed state where the bottom level is raised and aligned with the middle level. (**D**) Image of the stage system in its fully compressed state where the lower two levels are raised and aligned with the top level.

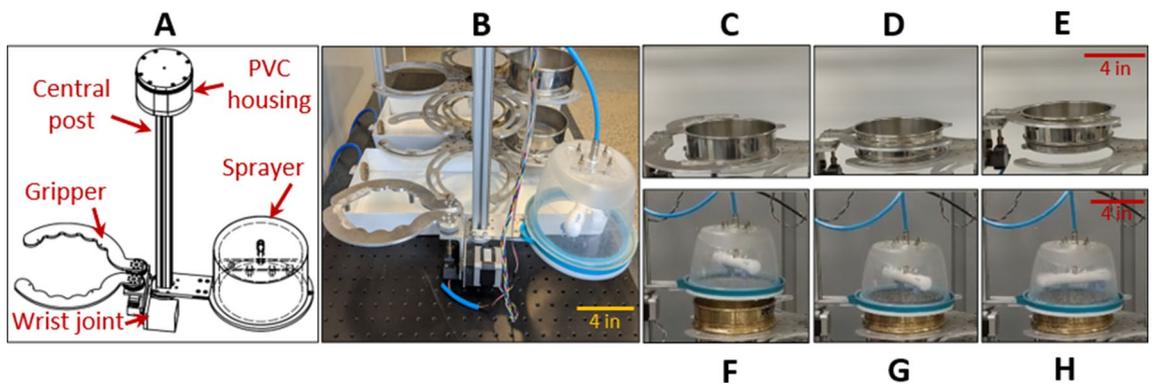

**Figure 3.** The gripper/washing system. (**A**) CAD schematic showing the gripper and water sprayer. (**B**) Image of constructed parts of this system and its relative position in the instrument. (**C**–**E**) Images of the gripper preparing to grab a sieve, holding the sieve, and lifting the sieve from its sieve slot in the stage system, respectively. (**F**–**H**) Images of the sprayer moving into position above a sieve, aligning with the sieve, and rinsing the sieve contents, respectively.

The 'gripper/washing system' consists of a robotic gripper armature and a water sprayer mounted about a central post (Figs. 1B, 3). The gripper has a wrist joint and two semicircular fingers. The wrist joint rotates the gripper vertically up to 180°, thereby inverting a sieve about a central pivot point. The two semicircular fingers are actuated by a servo to change the gripping state (i.e. open or closed) of the gripper while interacting with a sieve (i.e. to release or hold) as demonstrated in Movie S3. The water sprayer has a spray bar attached to a rotating push-to-connect hose connector. The sprayer is actuated by a solenoid valve to create sufficient water pressure for self-induced rotation and washing of sieve contents.

The 'grinding system' consists of a drill press, linear actuator, and grinding pad mounted above the stage system (Figs. 1C, 4). In the drill press, the linear travel of the quill is automated to lower the grinding pad onto the sieve screen during the grinding operation or to raise it clear of the sieves for the free rotation of the stage system. In conjunction with the drill press, a spray nozzle is mounted on the framework to rinse the grinding pad and facilitate the transfer of desired material to sieves at the lower levels. The grinding and washing operations to gently rupture the cysts to release their eggs are demonstrated in Movie S4.

The 'control electronics' consists of microcontroller boards, stepper motor drivers, and a multichannel relay module (Figs. 1D, S2). The microcontroller boards control the different operations of the instrument in response to instructions received from the user interface software. The stepper motor drivers control the stepper motors in the stage system. The multichannel relay module controls the solenoid valves that regulate the water flow. A separate relay automates the electronic switching of the drill press. The user interface software (Fig. 1E) is implemented as a Graphical User Interface (GUI) and displayed on a portable touchscreen display. The GUI serves as the central command station for the protocols for soil processing and extraction of nematode cysts and eggs (Fig. S4).

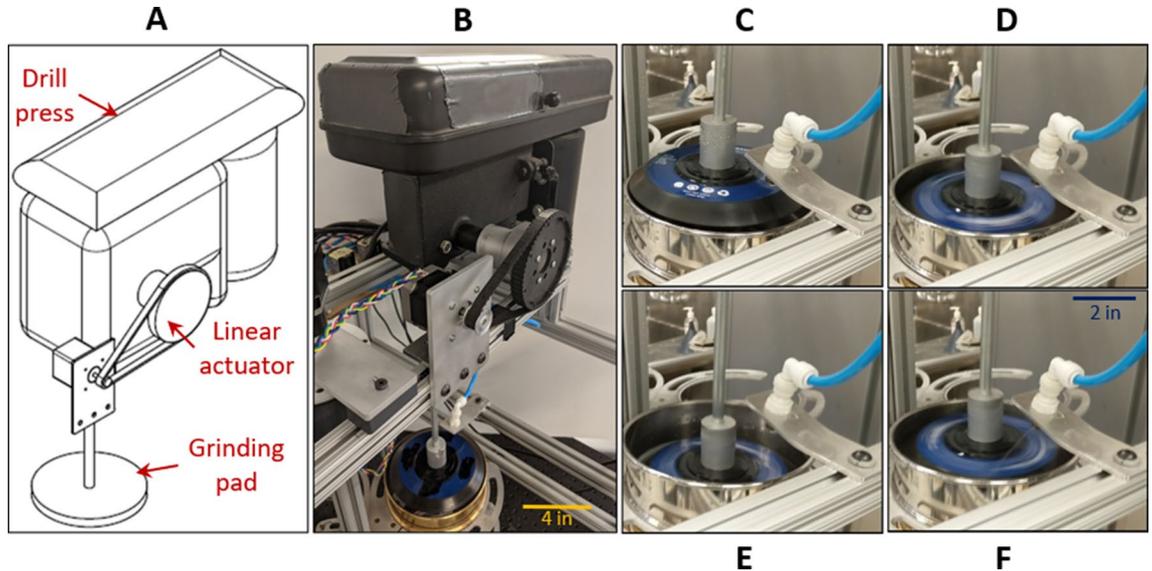

**Figure 4.** The grinding system. (**A**) CAD schematic showing the drill press, linear actuator, and grinding pad. (**B**) Image of constructed parts of this system and its relative position in the instrument. (**C**–**F**) Images of the grinding pad in its neutral position, being turned on and lowered within the sieve, gently grinding material on the sieve screen, and being washed by a spray nozzle in position above the sieve screen.

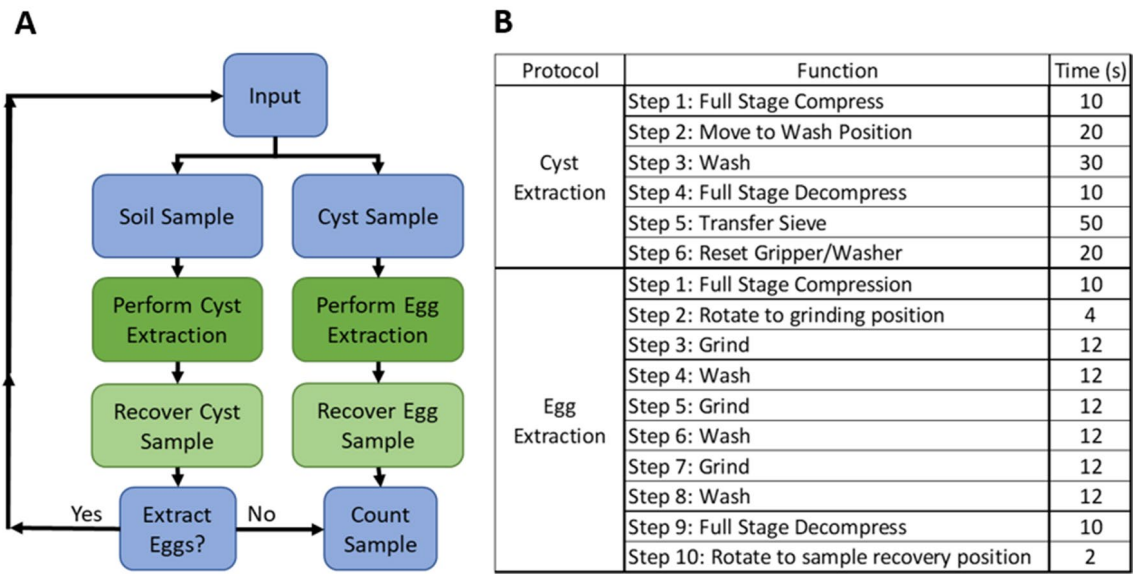

**Figure 5.** Process flow and extraction protocols. (**A**) Flowchart depicts the overall process flow for the robotic instrument. The user selects an input type (soil sample for cyst extraction or cyst sample for egg extraction). The appropriate extraction protocol is chosen and performed. The resultant sample is recovered, and the count of cysts or eggs is obtained. (**B**) List of the robotic functions implemented by the two protocols for cyst and egg extraction, along with the time duration of each function.

**Protocol for nematode cyst extraction.** The overall process flow is illustrated in Fig. 5A. The sample preparation starts by adding 100 cc of soil to two quarts of water. The soil suspension is mixed for 15 s and left to settle for 15 s. Four mesh sizes of sieves are used: #20 sieve with 850 μm pore size, #60 sieve with 250 μm pore size, #200 sieve with 75 μm pore size, and #500 sieve with 25 μm pore size. The user then selects the cyst extraction or egg extraction protocol and its embedded functions as listed in Fig. 5B. A dry run of the robotic instrument is shown in Movie S5.

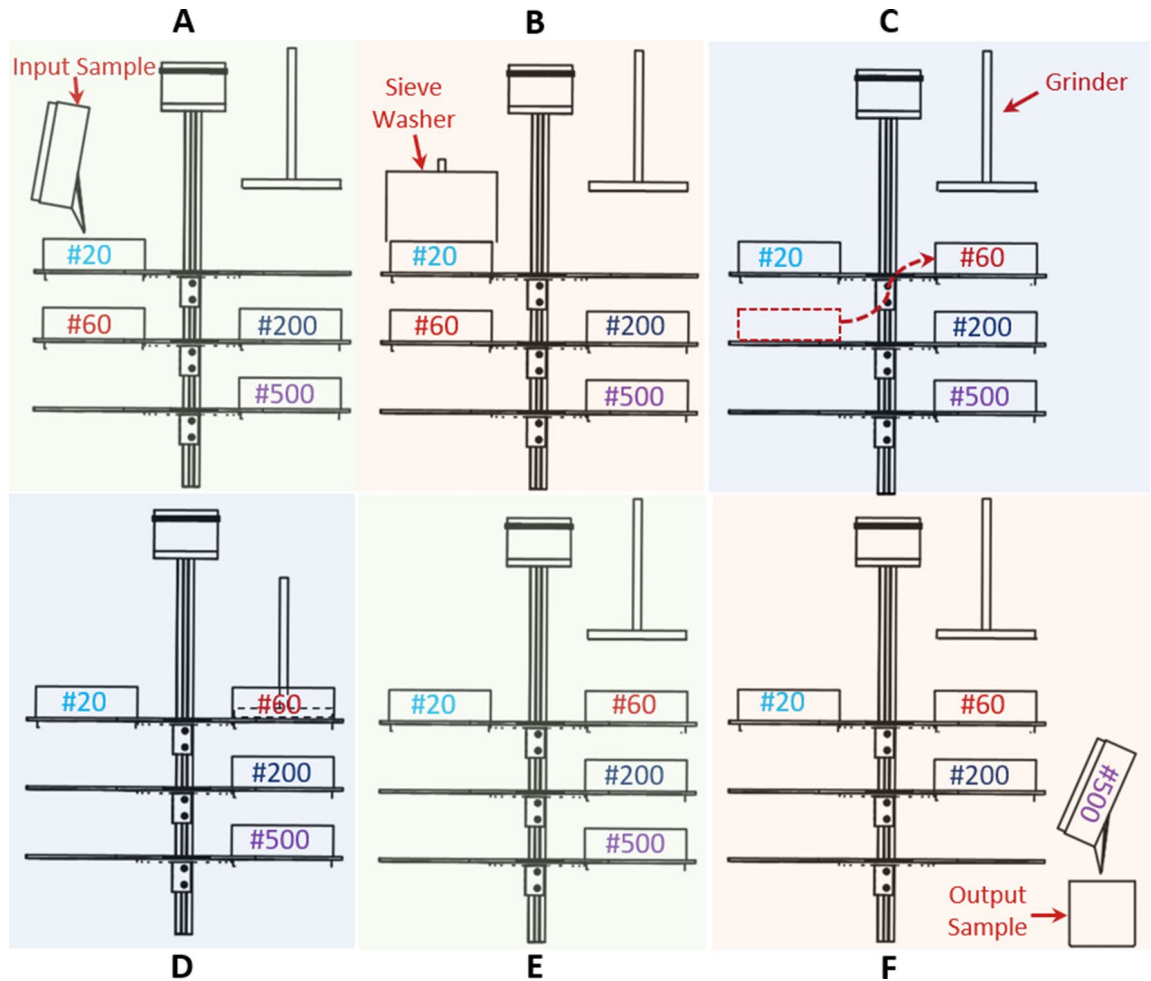

**Figure 6.** Manipulation of sieves during the extraction protocols. (**A**) The soil suspension is decanted through a #20 sieve seated above a #60 sieve. (**B**) The water sprayer moves into place above the #20 sieve to wash its contents and ensure the extracted cysts pass through to the #60 sieve. (**C**) The cyst-containing #60 sieve is transferred via the gripper onto the top stage level. (**D**) The grinding pad is brought into contact with the cyst-containing #60 sieve to rupture the cysts and release their eggs. (**E**) The grinder is raised to be free of the sieve and turned off. (**F**) The eggs are collected on the #500 sieve and transferred to a container to be counted later.

The protocol for nematode cyst extraction (Fig. 5B) is automatically executed by activating a set of functions through the GUI (Fig. S4). The relative arrangement of sieves is depicted in Fig. 6A–C. The soil suspension is decanted over the #20 sieve resting on the top stage level to trap large material such as root debris and different sized minerals (Fig. 6A). Below the #20 sieve, the middle stage level holds the #60 sieve which traps the cysts and other similarly sized material. The stage system is rotated into alignment with the gripper/washing system and is actuated into a fully compressed state to vertically stack the #20 and #60 sieves. The washing system is moved to cover the #20 sieve and the water sprayer is turned on to wash the decanted soil sample for 30 s (Fig. 6B). The stage system is actuated back to its fully uncompressed state, providing the gripper access to manipulate the sieves. The washing system is raised and rotated away from the stage system. The gripper moves into position to grab the #60 sieve and place it on the top stage level above the #200 and #500 sieves (Fig. 6C) in preparation for the egg extraction protocol. The entire cyst extraction process (Fig. 6A–C) takes around 2 min and 20 s.

**Protocol of nematode egg extraction.** The protocol for nematode egg extraction (Fig. 5B) from cysts is automatically executed by activating a set of functions through the GUI (Fig. S4). The relative arrangement of sieves is depicted in Fig. 6D–F. Initially, the stage system is rotated to align the #60 sieve with the grinding pad. The stage system is actuated into a fully compressed state to vertically stack the #60, #200 and #500 sieves. The grinding pad is lowered until it rests 1-inch above the #60 sieve mesh (Fig. 6D). The drill press is turned on to rotate the grinding pad at a high speed (~ 500 rpm). The grinding pad is simultaneously lowered until it is in contact with the sieve mesh. The fast-rotating motion of the grinding pad on the sieve mesh helps to rupture the cysts and dislodge their eggs. The grinding step is continued for 10 s. Then grinding pad is raised 1-inch above and sprayed with a water nozzle for 10 s before returning to contact the sieve mesh. The washing step facilitates the transfer of smaller particles through the #200 sieve and onto the #500 sieve (where the eggs and other similarly sized debris are collected). The foregoing steps of grinding and washing are repeated three times as demonstrated in Movie S4. Then the drill press is turned off and the grinding pad is vertically raised out of the #60 sieve (Fig. 6E). The output sample having the extracted nematode eggs is transferred from the #500 sieve into a container (Fig. 6F) for subsequent viewing and counting under a microscope. The entire egg extraction process (Fig. 6D–F) takes around 1 min and 38 s.

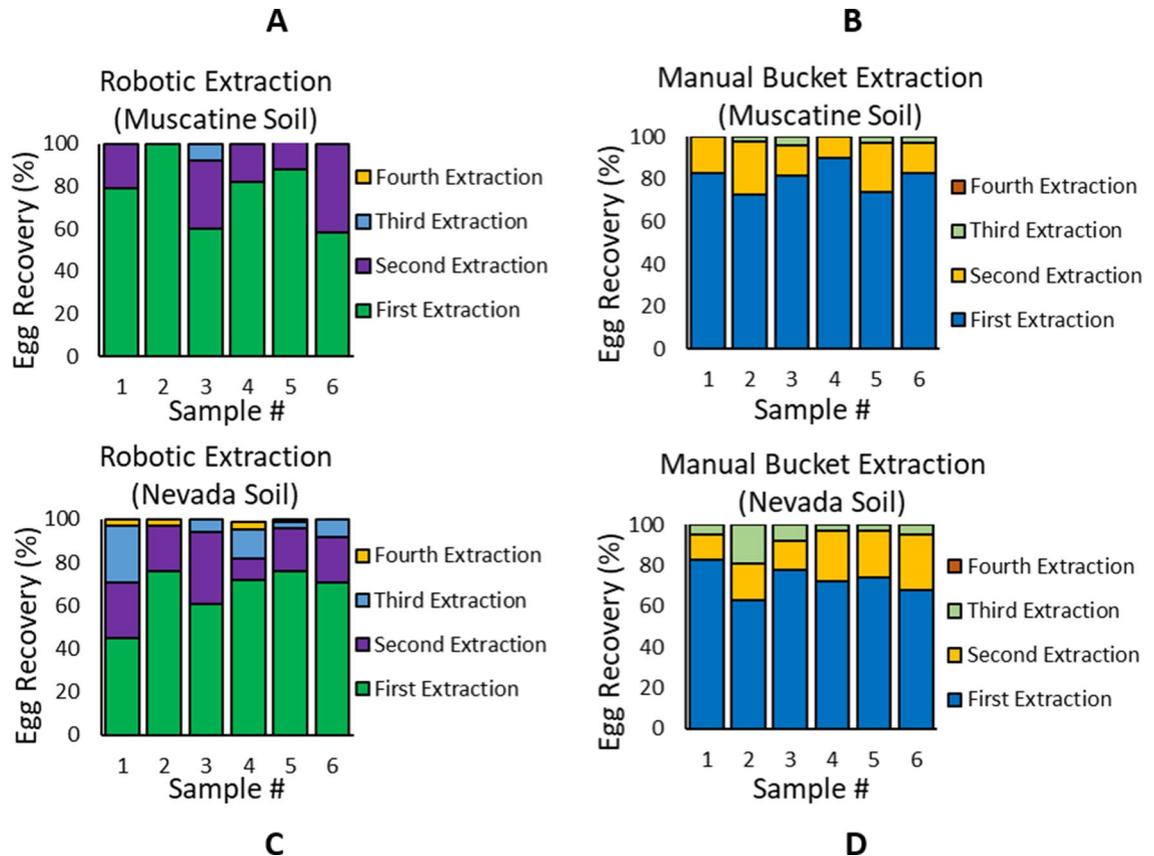

**Figure 7.** Performance of the robotic instrument for nematode egg extraction from field soil. (**A**,**C**) Plots of the nematode egg recovery percentage using the robotic instrument. (**B**,**D**) Plots of the nematode egg recovery percentage using the manual bucket method (i.e. wet-sieving technique). The field soil was obtained from fields near the cities of Muscatine and Nevada in Iowa. Each soil sample was processed to extinction (until 0 eggs were recovered) through four successive iterations. Six separate soil samples were tested for each case (n = 6 samples × 2 soil types × 2 extraction methods). For Muscatine soil samples, the first extraction iteration yielded an egg recovery percentage of 77.8 ± 14.8% (n = 6 samples) for the robotic instrument and 80.8 ± 5.8% (n = 6 samples) for the manual bucket method. For the Nevada soil samples, the first extraction iteration yielded an egg recovery percentage of 66.8 ± 11% (n = 6 samples) for the robotic instrument and 73.0 ± 6.5% (n = 6 samples) for the manual bucket method. In all experiments, more than 94% of nematode eggs were retrieved within the first two iterations of extraction.

*Performance of nematode egg extraction protocol.* We evaluated the performance of the robotic instrument for cyst nematode egg extraction in comparison with the conventional wet-sieving technique (commonly called the manual bucket method). Field soil was collected from soybean fields near two Iowa cities: Muscatine and Nevada, USA. We mixed the soil from collected from each field well and then processed multiple samples of each soil to extinction (until 0 eggs were recovered) using the robotic method (n = 12 with 6 Muscatine samples and 6 Nevada samples) and the wet-sieving technique (n = 12 with 6 Muscatine samples and 6 Nevada samples). Each soil sample underwent 4 successive iterations of extraction, wherein the left-over soil in the bucket after one iteration was re-suspended in water and processed in the next iteration (following the steps shown in Fig. 6A–F). The eggs were visually observed under a microscope and counted using a nematode-counting slide. The egg counts were used to calculate the percent recovery of eggs per extraction run as plotted in Fig. 7. For Muscatine soil samples, the egg recovery percentage in the first iteration of extraction was 77.8 ± 14.8% (n = 6 samples) for the robotic instrument and 80.8 ± 5.8% (n = 6 samples) for the wet-sieving technique. For the Nevada soil samples, the egg recovery percentage in the first iteration of extraction was 66.8 ± 11% (n = 6 samples) for the robotic instrument and 73.0 ± 6.5% (n = 6 samples) for the wet-sieving technique. In all cases, the majority of nematode eggs (i.e. greater than 94%) were retrieved in the first two iterations of extraction.

## Discussion

Knowing the population density (or number) of SCN cysts or eggs through routine soil analysis of fields, especially in the fall after the crops have been harvested, is widely acknowledged as a key first step towards successful long-term management of the pest[12,14,26,29]. Currently there is a nation-wide campaign called the SCN Coalition, funded by soybean farmers and the soybean industry, and their slogan is *Take the test, Beat the pest. What's your Number?* to create awareness about testing fields to determine SCN population densities[18]. The risks of not knowing the presence and population density of SCN in a timely manner can be detrimental to future crop yields[7,9]. For instance, if infestations are discovered late when the SCN population densities are high and very damaging (egg population density > 12,000 per 100 cc. of soil), it is difficult to reduce the levels of nematodes to manageable numbers again[12,32,33]. In unmanaged fields, SCN population densities can potentially increase 10- to 30-fold per growing season, and SCN infection of soybeans increases the vulnerability of the crop to other diseases (such as soybean sudden death syndrome)[34]. To this end, our robotic instrument was developed to simplify the extraction of nematode cysts and eggs from soil samples, which typically requires considerable time, labor, and patience of trained personnel when conventional extraction methods are used. Information gathered about the SCN population densities in their fields can reveal to soybean farmers if SCN levels are being kept consistently low (egg population density < 2000 per 100 cc. of soil) through use of standard management strategies of growing SCN-resistant soybean varieties in rotation with non-host crops[35–42]. Efforts also are underway by the agricultural industry to develop new resistant soybean varieties, nematode-protectant seed treatments, beneficial microbes, and transgenic resistance[43–46]. While the effectiveness of different soybean disease management strategies continues to be investigated, the soil processing techniques to determine cyst and egg population densities have largely remained unaltered over the past four decades[20–23,47,48].

Today, automation is being embraced by various aspects of farming such as for sampling, seeding, planting, harvesting, watering and irrigation, spraying, weeding, soil nutrient analysis, seed quality assessment, and remote surveillance of crop diseases[49–51]. Automating the critical process of extracting cysts and eggs of nematodes from soil to estimate their population densities brings benefits to SCN soil diagnostics. For instance, the need and cost to employ people with specialized training in the extraction and counting of plant-parasitic nematodes is greatly reduced. In general, automation also lends to reduced human bias and greater consistency in performance and time[52,53]. This is anticipated when a large number of soil samples need to be processed on a daily basis (i.e. > 20 samples per day in a soil diagnostic clinic)[26,27]. Furthermore, the soil processing time is standardized in the robotic instrument where one sample (100 cc of soil) takes approximately 4 min (cyst extraction: 2 min 20 s, egg extraction: 1 min 38 s) (Fig. 5B, Movie S5). In comparison, the soil processing times for the conventional manual sample extraction take between 7 and 10 min.

The robotic instrument features a flexible and modular design where the overall architecture can be subdivided into smaller systems and treated independently of each other (i.e. stage, gripper/washing, grinding/spraying, control electronics, and GUI)[54,55]. The user has the flexibility to replace sieves of different pore sizes, alter the movement of individual components, add more parts to the framework, and modify the sequence of operations. By using sieves of different pore sizes, a user can adapt the same instrument to other soil-borne organisms. The table-top size, low-cost hardware modules, light-weight structural framework, and hands-free operations controlled by a touchscreen GUI help improve the portability and user-friendliness of the instrument. The low voltage electronics and motors allow the platform to be run anywhere there is access to a standard 120 VAC outlet. Water can be supplied by a municipal water source with a suitable faucet diverter valve attachment.

There is further scope to utilize the robotic instrument. It can be used to extract other plant-parasitic nematode cysts and eggs as well as non-cyst-forming nematodes and large soil-borne fungal spores (like those of mycorrhizal fungi), weed seeds and many other soil fauna and flora[19,56–59]. There are over 4100 plant-parasitic nematodes known so far which cause over $100 billion dollars of crop damage every year[60–62]. Among these, there are eight different genera of cyst nematode with dozens of species known to exist throughout the world, and species in two genera cause economically significant yield loss to crops[63]. The robotic instrument would be able to be used to extract nematode cysts (egg-filled dead females) of all of these cyst nematode species from soil and their eggs from the cysts. In this regard, we do not anticipate any major alterations to our system other than changing the sieves and time durations for some operations, as the wet-sieving and decanting extraction procedure is similar for the other biological particles mentioned. There are further opportunities to couple the extraction and counting steps where the processed sample from the robotic instrument is automatically transferred into high-throughput nematode egg counting platforms[24].

It may take some time and effort for potential users to realize the reduced cost and additional benefits described herein from adopting the new technology relative to using the traditional wet sieving extraction method[25,26]. Also, it is important to emphasize that the robotic instrument only simplifies the process of determining the population densities of cyst nematode eggs from soil samples collected from fields in which soybeans were grown or will be grown. Its use does not guarantee or predict a reduced nematode infection rate or disease suppression or yield loss. To mitigate yield losses from SCN, farmers and those who advise them must collect soil samples to monitor SCN population densities on a regular basis (such as before or after every third soybean crop) and must also implement a sustained and coordinated effort to use every available control option in an integrated SCN management plan.

## Materials and methods

The framework (24 inch × 34 inch × 26 inch) of the robotic instrument was constructed from 80/20 Inc, 10 Series aluminum extrusions. The custom components were designed with the Autodesk Software Suite and machined from solid material (e.g. aluminum plates, acrylic, or PVC) using a CNC mill. PVC material was used to make the housings for the linear stepper motors used in the gripper and stage system. Acrylic material (thickness = 0.252 inch) was used as the mounting plate for the control board electronics. The remaining components were cut from aluminum plates (thickness = 0.375–0.5 inch). These materials were purchased from McMaster–Carr.

During the operation of the robotic instrument, the water supply is provided by a faucet connected to the municipal water source and monitored by a flow meter (Digiten, Quick Connect 0.3–10 L/min Water Hall effect Flow Sensor). The rotational speed of the paddle and the sensor pulse rate are monitored and correlated with the water flow rate from faucet. Thereafter, the water supply is fed into a sprayer having a rotating push-to-connect hose attachment that affixes to the spray bar. The spray bar is custom designed to have two asymmetrically-located downward-pointing spray nozzles and two downward-angled spray nozzles. The water pressure in the angled spray nozzles produce a net torque on the bar to propel its rotation while the downward spray nozzles ensure maximum coverage of the sieve surface area. The discarded water is eventually collected in a system consisting of a drainage pan connected to a 5-gallon plastic bucket (Fig. S3). The drainage pan is designed to slide around the base of the stage system, and is fabricated using a vacuum former and plywood mold.

The control board integrates two microcontroller boards (Arduino Unos, Adafruit Industries) which control six Pololu A4988 stepper motors drivers, a servo motor (Futaba S3305), and an 8-channel SunFounder relay board module—all of which were purchased from the Robot Shop. The microcontroller boards are powered via their serial connections to a Raspberry Pi 3 Model B, while the relay module is powered by a 12 VAC connection. One microcontroller board is dedicated to communicating with the sensors, actuating the relay module, and manipulating the servo motor. The second microcontroller board controls and monitors the functioning of the robotic instrument. The motor driver board and its circuitry are powered by a 12 VAC, 2 A wall supply, which then powers the stepper motors through the motor driver modules. The servo motor and rest of the electronics on the control board are powered by regulating this 12 VAC supply down to 5 VAC. The motor selection is achieved by means of a multiplexer and inverter. Together, these chips enable the respective motor driver modules in response to the pulse width modulated signal generated by the associated microcontroller board. To control the system communication, a GUI is developed for a touchscreen-compatible Raspberry Pi single board computer (Fig. S4). Informed consent was obtained from Christopher Legner (first author of this submission) to publish the image (Supplementary Video 5) in an online open access publication.

### Data availability
All data are presented in the paper and/or in the Supplementary Materials. Please contact S.P. for additional information.

### Acknowledgements
We are grateful to David Soh (ISU Department of Plant Pathology and Microbiology) for extracting SCN cysts and eggs from soil samples manually and providing the manual egg counts. We are also thankful to Lee Harker (ISU Department of Electrical and Computer Engineering) for his help and support in the mechanical construction of the instrument. The NSF-sponsored ISU I-Corps program provided advice for commercialization. A patent application on the robotic instrument has been awarded through the United States Patent and Trademark Office (US Patent No. 10,900,877 issued on January 26, 2021, Patent Title: Methods, Apparatus, and Systems to Extract and Quantify Minute Objects From Soil or Feces, Including Plant-Parasitic Nematode Pests and Their Eggs in Soil).

### Author contributions
All authors planned the experiments, contributed to the analysis of data and discussion. C.M.L. constructed, assembled, and tested the robotic instrument. All authors wrote and revised the paper.

### Funding
This work was partially supported by the U.S. National Science Foundation (NSF IDBR-1556370) and United States Department of Agriculture (2020–67021-31964) to S.P. and G.T.

### Competing interests
The authors declare no competing interests.

### Additional information
**Supplementary Information** The online version contains supplementary material available at https://doi.org/10.1038/s41598-021-82261-w.

**Correspondence** and requests for materials should be addressed to S.P.

**Reprints and permissions information** is available at www.nature.com/reprints.

**Publisher's note** Springer Nature remains neutral with regard to jurisdictional claims in published maps and institutional affiliations.